\documentclass{article}

\usepackage{arxiv}

\usepackage[utf8]{inputenc} % allow utf-8 input
\usepackage[T1]{fontenc}    % use 8-bit T1 fonts
\usepackage{hyperref}       % hyperlinks
\usepackage{url}            % simple URL typesetting
\usepackage{booktabs}       % professional-quality tables
\usepackage{amsfonts}       % blackboard math symbols
\usepackage{nicefrac}       % compact symbols for 1/2, etc.
\usepackage{microtype}      % microtypography
\usepackage{lipsum}
\usepackage{graphicx}
\graphicspath{ {./images/} }

\usepackage{lineno,hyperref}

\usepackage{booktabs} % For formal tables
\usepackage{framed}
\usepackage{graphicx}
\usepackage{amssymb}
\usepackage{tabularx}
 \usepackage{graphicx}
 \usepackage{url}
 \usepackage{longtable}

\usepackage{textgreek}  
\usepackage{microtype}
\usepackage{textcomp}
\usepackage[caption=false]{subfig}
\usepackage{flushend}
\usepackage{csquotes}
\usepackage{booktabs} % For formal tables
\usepackage{framed}
\usepackage{graphicx}
\usepackage{amssymb}
\usepackage{graphicx}
\usepackage{url}
\usepackage{multirow}
\usepackage{enumitem}
\usepackage{hyperref}
\usepackage{lscape}
\usepackage[normalem]{ulem}
\useunder{\uline}{\ul}{}
\usepackage{xcolor}
\definecolor{lightgray}{gray}{0.9}
 
\usepackage{siunitx}  
\usepackage{textgreek}  
\usepackage{wasysym}
\usepackage{makecell}

 \usepackage{adjustbox}
  \usepackage{multirow}
 \usepackage[normalem]{ulem}
 
\title{Pandemic Pedagogy: Evaluating Remote Education Strategies during COVID-19}

\author{
Daniel Russo \\
  Department of Computer Science\\
 Aalborg University\\
  2450 Copenhagen \\
  \texttt{daniel.russo@cs.aau.dk} \\
}

\begin{document}
\maketitle
\begin{abstract}
The COVID-19 pandemic precipitated an abrupt shift in the educational landscape, compelling universities to transition from in-person to online instruction. This sudden shift left many university instructors grappling with the intricacies of remote teaching. Now, with the pandemic behind us, we present a retrospective study aimed at understanding and evaluating the remote teaching practices employed during that period. Drawing from a cross-sectional analysis of 300 computer science students who underwent a full year of online education during the lockdown, our findings indicate that while remote teaching practices moderately influenced students' learning outcomes, they had a pronounced positive impact on student satisfaction. Remarkably, these outcomes were consistent across various demographics, including country, gender, and educational level. As we reflect on the lessons from this global event, this research offers evidence-based recommendations that could inform educational strategies in unwelcoming future scenarios of a similar nature, ensuring both student satisfaction and effective learning outcomes in online settings. 
\end{abstract}


\section{Introduction}
\label{sec:introduction}

The COVID-19 pandemic has emerged as a dual challenge: a significant public health crisis and a disruptor of traditional norms and practices across multiple sectors. The computer science community has felt this impact profoundly~\cite{russo2020predictors,russo2023understanding}. Software professionals, akin to many professionals, faced a seismic shift in their daily routines as remote work became the new norm, influencing their job satisfaction and productivity levels~\cite{russo2021daily,russo2023satisfaction}.
Simultaneously, the realm of computer science education underwent a transformative phase, as underscored by Riese \& Kann~\cite{riese2021computer}.  The sudden closure of university campuses has thrust education into a virtual paradigm, causing both educators and students to adapt to an entirely new mode of learning and interaction.
This sudden shift from physical to online teaching has placed university instructors in a particularly challenging position. The traditional methods and pedagogies, once effective within the confines of a classroom, have been challenged by the virtual space, where engagement, interaction, and assessment take on different dimensions~\cite{conrad2022student}. Instructors, often with limited experience and resources in online teaching, found themselves having to reinvent their approach, all the while maintaining educational standards and student engagement.
The complexity of this transition is further exacerbated by the heterogeneity of the student population, marked by diverse backgrounds, learning styles, and access to technology. The question of how to tailor remote teaching to suit various needs, without losing the essence of the subject matter, has become a central concern. Moreover, the psychological and social impacts of remote learning, including feelings of isolation and lack of motivation, have added layers of complexity to this already intricate landscape~\cite{decoito2022online}. 

While the COVID-19 pandemic has prompted a plethora of research into the challenges and dynamics of online teaching, the specific interplay between pedagogical practices, student experiences, and contextual factors within the domain of computer science education remains underexplored. Walker \& Koralesky~\cite{walker2021student} highlighted the shifts in student and instructor perceptions of engagement after the rapid transition to online teaching, emphasizing the general challenges faced across disciplines. Similarly, Conrad et al.~\cite{conrad2022student} investigated the factors influencing students' satisfaction with online learning environments, revealing that information overload and technical skill requirements can negatively impact the online learning experience. However, these studies, while insightful, often adopt a broad perspective, encompassing a range of disciplines and not delving deeply into the unique challenges and opportunities presented within computer science education. Decoito and Estaiteyeh~\cite{decoito2022online} explored STEM teachers' curriculum and assessment practices during the pandemic, but the specific nuances of computer science education, with its distinct pedagogical needs and challenges, were not the primary focus. While Rossi et al.~\cite{rossi2021active} emphasized the benefits of active learning tools in higher education during the pandemic, their study did not specifically target the computer science discipline. Thus, there is a pressing need for research that delves into the intricacies of remote teaching practices tailored to computer science education, bridging this evident gap in the literature.

Therefore, this paper aims to shed light on the most effective remote teaching practices during the COVID-19 pandemic, focusing specifically on computer science students. Through a cross-sectional study involving \(N=300\) participants exposed to one year of remote education, we explore the multifaceted dynamics of remote teaching. Our findings contribute nuanced insights into how remote teaching practices influence learning outcomes and satisfaction, revealing that these practices have little effect on learning but significantly enhance satisfaction. 
Interestingly, we observe a remarkable homogeneity in these effects, transcending variables such as country, gender, and education level. By examining the nuanced interplay between pedagogical practices, student experiences, and contextual factors, this paper fills a critical gap in the literature, providing valuable guidance for educators, administrators, and policymakers navigating the evolving landscape of remote education in the field of computer science.

The structure of this paper is organized as follows: In Section~\ref{sec:related}, we provide a comprehensive review of the relevant literature, drawing connections between various research areas that underpin our study. Section~\ref{sec:ResearchModel} delves into our research framework and outlines the hypotheses under investigation. Section~\ref{sec:design} details the research instrument, PLS-SEM, and describes the data collection process along with the demographics of the participants. The complete analysis is presented in Section~\ref{sec:analysis}. Our findings are then contextualized and discussed in Section~\ref{sec:discussion}, emphasizing their implications for remote education as also the limitations of this study. Finally, Section~\ref{sec:Conclusions} offers concluding thoughts and future works directions.

\section{Related Work}
\label{sec:related}

The COVID-19 pandemic has had a profound impact on education, leading to a sudden shift from traditional classroom teaching to online learning. Researchers have approached this unprecedented transition from various angles, exploring instructional strategies, student perceptions, subject-specific insights, institutional responses, and the broader challenges and opportunities of remote learning. The following sections discuss the most relevant themes in the literature.

\subsection{Instructional Strategies and Teaching Methodologies}
The theme of instructional strategies and teaching methodologies in remote learning is a complex one, enriched by various contributions. Nguyen et al.~\cite{nguyen2021insights} emphasize the student's role, highlighting the importance of active learning strategies and technological engagement. This is contrasted by Mahmood's~\cite{mahmood2021instructional} examination of how educators can adapt traditional teaching practices for online use, focusing on strategies like synchronous and asynchronous teaching.

Orlov et al.~\cite{orlov2021learning} add a theoretical perspective, arguing that the way teaching is conducted is more critical than the students' demographics. They explore various instructional strategies, such as flipped classrooms, resonating with both Nguyen et al. and Mahmood but offering a broader theoretical framework.

Pandey et al.~\cite{pandey2021covid} take a practical approach, providing a comprehensive framework for delivering online classes. This includes guidelines, tools, and strategies that encapsulate elements from the other papers, forming a step-by-step guide that bridges theory and practice.

Together, these works present a multifaceted view of instructional strategies in remote teaching. While all recognize the importance of engagement, adaptability, and technology, they offer unique insights from the student's experience to practical adaptation and theoretical frameworks. This collective perspective provides valuable guidance for navigating the complexities of remote teaching during the COVID-19 pandemic.

\subsection{Student Perceptions and Satisfaction}

The exploration of student perceptions and satisfaction in remote learning during the COVID-19 pandemic offers valuable insights into the successes and challenges of this unprecedented shift in education. Two papers in particular contribute to this theme, each focusing on different aspects of student perceptions.

Conrad et al.~\cite{conrad2022student} delve into the multifaceted nature of student perceptions, examining how students' beliefs about online learning difficulty influence their overall satisfaction. Their research reveals that factors such as technological proficiency, instructor responsiveness, and course design can significantly affect student satisfaction levels. By exploring these relationships, Conrad et al. provide a comprehensive view of what influences student perceptions and satisfaction in remote learning, extending beyond mere difficulty.

In contrast, Asgari et al.~\cite{asgari2021observational} conduct an observational study that focuses on specific challenges in engineering education. Their research uncovers technological barriers and motivational issues that have particular resonance in the context of engineering. By pinpointing these challenges, Asgari et al. highlight the nuanced differences that subject matter can introduce into remote learning experiences.

Together, these studies provide a well-rounded view of student perceptions and satisfaction. While Conrad et al. offer a broad exploration of influencing factors, Asgari et al. provide a subject-specific look at the unique challenges of remote teaching in engineering. The interplay between these two perspectives forms a richer understanding of the student experience in remote learning, recognizing both general trends and subject-specific nuances. This combination of insights helps to shed light on the ways educators can enhance student satisfaction and overcome challenges in various educational contexts during the pandemic.

\subsection{Institutional Responses and Transition to E-Learning}

The transition to online teaching during the COVID-19 pandemic has required not only individual adaptation by teachers but also broader institutional responses. Several papers contribute to this theme, shedding light on various aspects of this complex transition.

DeCoito \& Estaiteyeh~\cite{decoito2022online} focus on the specific field of STEM education, examining how teachers have adapted their curriculum and assessment practices to the online environment. They emphasize the integration of technology and the innovative approaches that teachers have developed to engage students in scientific inquiry remotely. Their work provides insights into the subject-specific challenges of teaching STEM disciplines online and illustrates how educators have navigated these challenges.

Turnbull et al.~\cite{turnbull2021transitioning} take a broader institutional perspective, exploring how higher education institutions have responded to the challenge of transitioning to e-learning. They analyze institutional strategies, policies, and support systems, identifying both successes and areas for improvement. Their findings reveal the vital role that institutional leadership and support play in enabling a successful transition, emphasizing the need for clear communication, faculty training, and technological infrastructure.

Complementing this institutional view, Cutri et al.~\cite{cutri2020faculty} investigate faculty readiness for online crisis teaching. They measure various factors such as training, support, and adaptability, providing a nuanced understanding of faculty experiences and needs. Their research underscores the importance of timely training and ongoing support to ensure that faculty are not only ready but also confident in their ability to teach online.

Together, these papers provide a comprehensive view of the institutional responses and transitions to e-learning. They reveal the multifaceted nature of this transition, encompassing subject-specific curriculum adaptation, institutional strategies and policies, and faculty readiness and support. While each paper focuses on different aspects, they collectively highlight the interconnectedness of these elements, emphasizing the need for a coordinated and supportive approach at all levels. This multifaceted perspective offers valuable insights for educational leaders, policymakers, and practitioners, guiding them in the complex task of transitioning to online teaching in a crisis situation.

\subsection{Opportunities and Challenges of Remote Learning}

The transition to remote learning during the COVID-19 pandemic has presented both significant challenges and unique opportunities. Several papers contribute diverse perspectives to this theme, providing a multifaceted examination of remote learning across different contexts and educational levels.

Adedoyin \& Soykan~\cite{adedoyin2023covid} provide a broad analysis of the challenges and opportunities of online learning. They identify key challenges such as technological barriers, accessibility issues, and engagement difficulties, while also highlighting the opportunities for flexible and innovative approaches. Their research emphasizes the importance of adaptability and creativity in remote learning, shedding light on how these can turn challenges into opportunities for enhanced learning experiences.

Oliveira et al.~\cite{oliveira2021exploratory} carry out an exploratory study that focuses on the need for effective communication, support, and guidelines in emergency remote education. They pinpoint critical success factors and challenges in higher education institutions' response to the pandemic, emphasizing the importance of clear communication, robust support systems, and well-defined guidelines. Their insights contribute to a broader understanding of how institutions can create conducive remote learning environments.

Adding to this discussion, Aslan et al.~\cite{aslan2021teachers} examine teachers' views on middle school curriculum adaptation for distance education. They emphasize the importance of collaboration among teachers and the need for comprehensive guidelines to ensure effective curriculum delivery. Their focus on middle school education adds a specific educational level's perspective, highlighting the unique challenges and opportunities at this stage.

Collectively, these studies provide a broader view of remote learning, encompassing various angles such as overall challenges and opportunities, institutional responses, communication strategies, and specific educational levels. They reveal a complex landscape where challenges such as technological barriers and engagement difficulties coexist with opportunities for innovation, collaboration, and flexibility. By exploring these facets, the papers offer valuable insights that can guide educators, administrators, and policymakers in leveraging the opportunities and overcoming the challenges of remote learning during a global crisis.

\subsection{Subject-Specific Insights in Software Engineering and Computer Science Education}

The COVID-19 pandemic has ushered in a period of rapid transformation in higher education, with unique challenges and opportunities emerging in the fields of software engineering and computer science education. Several studies delve into this subject, each providing a different angle on the transition to online learning.

Barr et al.~\cite{barr2020online} reflect on the experience of delivering an eight-week undergraduate Software Engineering program during the pandemic, emphasizing the possibility of successful online delivery with thoughtful adjustments. They highlight that there is no `one size fits all' approach, and by following established best practices such as breaking lectures into smaller `chunks', a pedagogically sound experience can be achieved. This perspective resonates with the findings of Iglesias-Pradas et al.~\cite{iglesias2021emergency}, who analyzed emergency remote teaching at the School of Telecommunication Engineering. They found that students' academic performance increased under emergency remote teaching, supporting the idea that organizational factors contribute to success, while class size and delivery modes did not significantly impact performance.

While Barr et al. and Iglesias-Pradas et al. focus on the adaptability and outcomes of online delivery in engineering contexts, Crick et al.~\cite{crick2020impact} take a broader view of the UK Computer Science Education community. Their large-scale survey revealed more positive attitudes towards online learning in computer science than in other disciplines, yet uncovered specific concerns about delivering core topics and formal assessments. This focus on discipline-specific challenges aligns with Barr et al.'s reflections on the unique needs of software engineering education.\\

Together, these studies paint a complex picture of the subject-specific challenges and successes experienced during the pandemic. They reveal a landscape where successful online delivery is possible but requires careful consideration of subject-specific needs and organizational readiness. By emphasizing flexibility, best practices, and recognition of the unique characteristics of each discipline, they contribute valuable lessons for educators, administrators, and policymakers navigating the ongoing evolution of online teaching and learning.

\section{Research Model and Hypotheses}
\label{sec:ResearchModel}

This section presents a structured exploration into the determinants of effective remote teaching during the COVID-19 era. Drawing from the constructivist learning theory and recent research, we posit a series of hypotheses to understand the impact of various pedagogical practices on student outcomes and satisfaction. Each subsequent subsection delves into the specifics of these hypotheses, providing a clear roadmap for our investigative approach.

\subsection{Constructivist Learning in the Digital Age and its Relevance to Remote Teaching during the COVID-19 pandemic}

The COVID-19 pandemic instigated a abrupt shift in the educational sector, forcing universities to transition from traditional classroom settings to online platforms. This shift change had numerous challenges for educators, who found themselves navigating the difficulties of remote teaching. Central to addressing these challenges is the constructivist learning theory, deeply rooted in the works of thinkers like Piaget~\cite{piaget1977} and Vygotsky~\cite{vygotsky1978}. This theory posits that learners actively construct knowledge through their interactions and experiences. In the context of remote education during the pandemic, understanding how students construct knowledge becomes even more critical, especially when considering the pivotal indicators of the efficacy of online education: \textit{learning outcomes} and \textit{student satisfaction}~\cite{alavi1995using,graham2001enhancing}.

Historically, constructivist learning in conventional classrooms has been characterized by hands-on activities, collaborative discussions, and problem-solving exercises. The digital realm, accelerated by the pandemic, offers tools and platforms that can emulate and even enhance these experiences. Online forums, virtual laboratories, and interactive simulations can serve as avenues for active learning, aligning with the constructivist paradigm.

However, the transition to online learning during the pandemic was not without its challenges. The absence of face-to-face interactions, potential feelings of isolation, and the vastness of online resources can be overwhelming for students. Given these challenges, educators have sought guidance from constructivist principles. This involves creating opportunities for active participation, fostering a sense of community, and providing structured guidance~\cite{jonassen1995supporting}.

Vygotsky's concept of the ``zone of proximal development''~\cite{vygotsky1978} emphasizes the importance of social interactions in learning. In the pandemic-induced remote learning context, this translates to synchronous online discussions, peer assessments, and group projects. Modern digital tools, such as video conferencing and shared document platforms, can help bridge the physical divide, fostering these interactions and ensuring continuity in learning.

The pandemic also underscored the potential for personalized online learning experiences. Adaptive platforms could tailor content to individual learners' needs, ensuring they consistently operated within their zone of proximal development. Such platforms, in alignment with constructivist principles, allowed learners to assimilate knowledge at their pace.

Now that the pandemic has passed, this research primarily aims to retrospectively delve into the factors influencing students' perceived learning outcomes and their satisfaction within the context of online university education during the lockdown. The lessons learned from this period offer valuable insights for future scenarios that might necessitate a similar shift. By understanding and leveraging these principles, educators can be better prepared to craft effective remote teaching strategies, ensuring optimal learning outcomes and heightened student satisfaction. In the subsequent subsection, drawing from existing literature, we will present a research model that delineates the elements impacting the outcomes of e-learning systems, preparing us for any future events of a similar nature.

\subsection{Research Hypotheses}

The sudden transition to remote teaching during the COVID-19 pandemic emphasized the need to understand effective online pedagogical practices. Drawing from the constructivist learning theory, this research seeks to identify the determinants of effective remote teaching. Our focus is on the factors influencing students' perceived learning outcomes and satisfaction during the lockdown.

\begin{figure}[ht!]
\centerline{\includegraphics[width=1\linewidth]{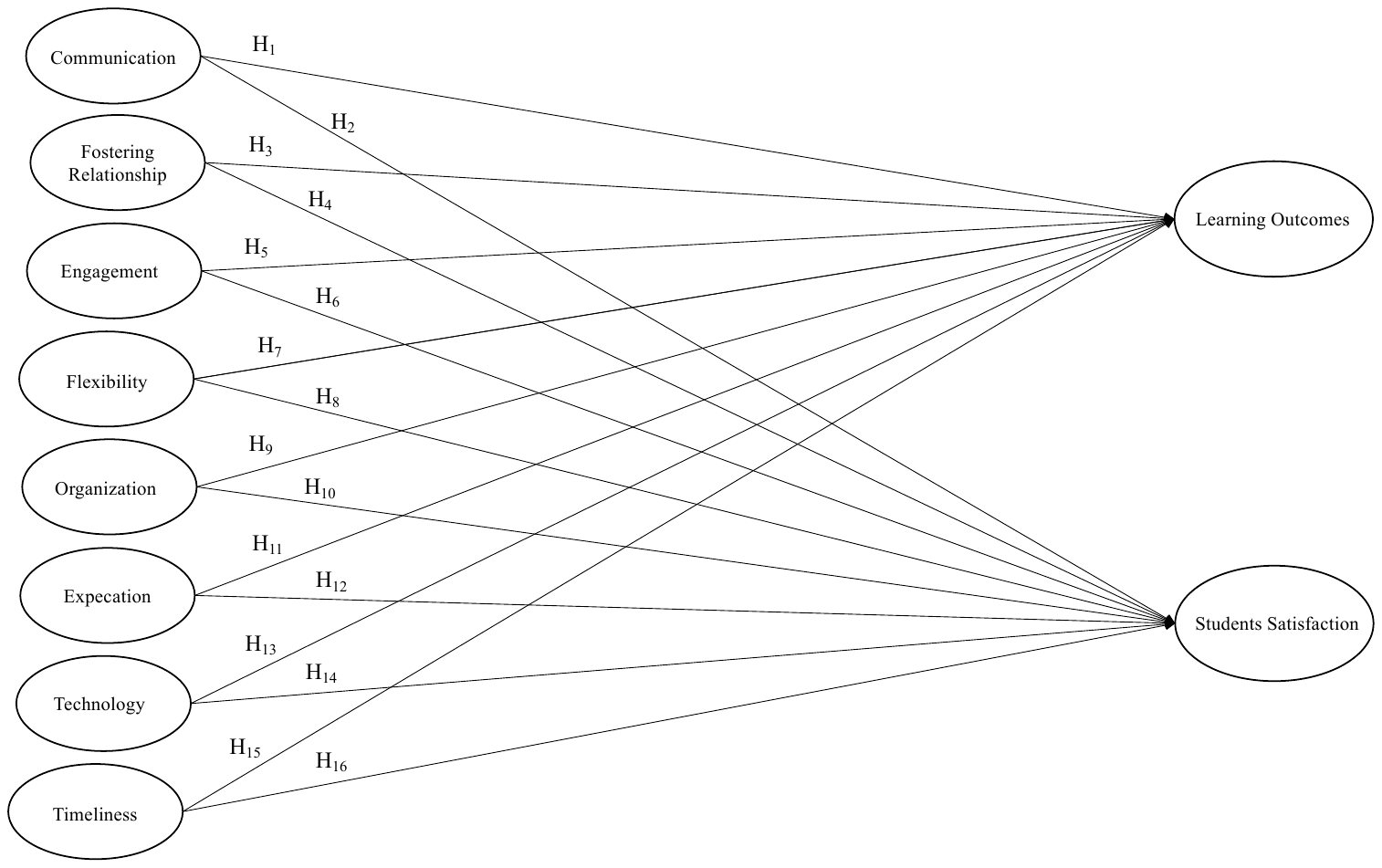}}
\caption{Theoretical model.}
\label{fig:TheoreticalModel}
\end{figure}

Our research hypotheses' independent variables are informed by the work of Craig and Card~\cite{bailey2009effective}. Their research provides a comprehensive exploration of the pedagogical practices that experienced instructors perceive as effective for online teaching. As we proceed through this subsection, we will introduce our research model, based on existing literature, that outlines the elements impacting e-learning education outcomes. This model is also visually represented in Figure \ref{fig:TheoreticalModel}.

Online teaching, especially when prompted by unforeseen events like the COVID-19 pandemic, presents both challenges and opportunities. The following subsections delve into various facets of effective online teaching, each accompanied by hypotheses that posit their influence on learning outcomes and student satisfaction.

\subsubsection{Communication in Online Teaching}

Effective communication is paramount in the online teaching environment. Unlike face-to-face settings, online platforms lack non-verbal cues, making the clarity of written communication crucial. Instructors must be particularly attentive to their communication styles and the words they use. Effective online communication requires both clarity and empathy. Given the importance of clear and empathetic communication in facilitating understanding and building trust:

\begin{center}
      \normalfont 
      \textit{H$_1$: Communication positively influences Learning Outcomes}
\end{center}

\begin{center}
      \normalfont 
      \textit{H$_2$: Communication positively influences Students Satisfaction}
\end{center}

\subsubsection{Fostering Relationships in the Virtual Classroom}

Building and fostering relationships in an online setting can be challenging, yet it remains a cornerstone of effective teaching. Instructors' empathy, passion for teaching, and genuine desire to help students succeed can bridge the virtual gap, creating a sense of community and belonging.

\begin{center}
      \normalfont 
      \textit{H$_3$: Fostering Relationships positively influence Learning Outcomes}
\end{center}

\begin{center}
      \normalfont 
      \textit{H$_4$: Fostering Relationships influence Students Satisfaction}
\end{center}

\subsubsection{Engagement in Online Learning}

Active engagement is essential for deep learning and retention. Instructors who prioritize engagement strategies, such as discussion boards and group projects, often report higher levels of student participation and satisfaction. Engaged students are more likely to be invested in their learning, leading to better outcomes.

\begin{center}
      \normalfont 
      \textit{H$_5$: Engagement positively influences Learning Outcomes}
\end{center}

\begin{center}
      \normalfont 
      \textit{H$_6$: Engagement positively influences Students Satisfaction}
\end{center}

\subsubsection{Flexibility in Online Instruction}

The dynamic nature of online teaching requires instructors to be adaptable and flexible. Whether navigating technical issues or adjusting to students' diverse needs, a flexible approach can enhance the learning experience and accommodate unforeseen challenges.

\begin{center}
      \normalfont 
      \textit{H$_7$: Flexibility positively influences Learning Outcomes}
\end{center}

\begin{center}
      \normalfont 
      \textit{H$_8$: Flexibility positively influences Students Satisfaction}
\end{center}

\subsubsection{Organization in Online Course Design}

A well-organized online course provides a clear roadmap for students, reducing confusion and enhancing their learning experience. Effective organization, from a user-friendly course website to clear guidelines, can significantly impact student success and satisfaction.

\begin{center}
      \normalfont 
      \textit{H$_9$: Organization positively influences Learning Outcomes}
\end{center}

\begin{center}
      \normalfont 
      \textit{H$_{10}$: Organization positively influences Students Satisfaction}
\end{center}

\subsubsection{Setting High Expectations}

Clearly defined expectations provide students with a clear understanding of what is required for success. By setting and communicating high expectations, instructors can motivate students to strive for excellence.

\begin{center}
      \normalfont 
      \textit{H$_{11}$: Expectation positively influences Learning Outcomes}
\end{center}

\begin{center}
      \normalfont 
      \textit{H$_{12}$: Expectation positively influences Students Satisfaction}
\end{center}

\subsubsection{Effective Utilization of Technology}

In the digital age, the effective integration of technology into teaching is not just beneficial—it is essential. Instructors who are adept at using technological tools can enhance the learning experience, making content more accessible and engaging.

\begin{center}
      \normalfont 
      \textit{H$_{13}$: Technology positively influences Learning Outcomes}
\end{center}

\begin{center}
      \normalfont 
      \textit{H$_{14}$: Technology positively influences Students Satisfaction}
\end{center}

\subsubsection{Timeliness in Online Teaching}

Prompt responses and feedback are crucial in online teaching. Timely interactions can alleviate student anxieties, provide clarity, and foster a sense of instructor presence, which can significantly impact student satisfaction and learning outcomes.

\begin{center}
      \normalfont 
      \textit{H$_{15}$: Timeliness positively influences Learning Outcomes}
\end{center}

\begin{center}
      \normalfont 
      \textit{H$_{16}$: Timeliness positively influences Students Satisfaction}
\end{center}

%In summary, each hypothesis, rooted in the constructivist learning theory and empirical research, emphasizes its potential impact on students' learning outcomes and satisfaction. These hypotheses provide a structured foundation for our subsequent analysis, aiming to shed light on the complexities of remote education during this unprecedented time.

\section{Research Design}
\label{sec:design}
This section outlines the methodological framework utilized in our study, focusing on the application of Partial Least Squares -- Structural Equation Modeling (PLS-SEM). Recognized for its ability to validate latent constructs through indicators, PLS-SEM is particularly relevant in the field of empirical software engineering. We detail the process of scale development, survey data collection procedures, and provide a comprehensive demographic breakdown of the participants, all of which contribute to the study's robustness. By capturing a holistic view of computer science students' experiences during the COVID-19 lockdown, this research design ensures a thorough and nuanced understanding of the subject matter. The methodology adheres to the ACM SIGSOFT Empirical Standards concerning Questionnaire Surveys, further enhancing the credibility and rigor of our approach~\cite{ralph2020acm}.

\subsection{Partial Least Squares -- Structural Equation Modeling (PLS-SEM)}

Partial Least Squares -- Structural Equation Modeling (PLS-SEM) represents a sophisticated statistical approach aimed at validating latent variables, also referred to as constructs, by means of multiple indicators~\cite{russo2021pls}. This technique is especially pertinent to the field of theoretical development studies and has found growing acceptance in the realm of empirical software engineering~\cite{russo2019soft,russo2023navigating,russo2023don,russo2021agile}. The multifunctional nature of PLS-SEM enables it to tackle several interrelated research questions within a comprehensive analysis, leading to its widespread utilization in related disciplines such as Management, Information Systems Research, and Organizational Behavior. According to scholars like Gefen et al.~\cite{gefen2000structural}, SEM is frequently applied to certify instruments and confirm the relationships between various constructs. The subsequent scrutiny and evaluation of the PLS-SEM model are conducted in alignment with the most current guidelines and best practices for research within the software engineering domain, as outlined by Russo \& Stol~\cite{russo2021pls}.

\subsubsection{Scale Development}

Our survey's design was informed and enriched by supplemental theoretical frameworks. We configured the survey by adapting instruments sourced from prior studies of Eom et al.~\cite{eom2016determinants} and Bailey \& Card~\cite{bailey2009effective}. A comprehensive summary of each construct, along with the references employed to define the questions, can be found in Table \ref{tab:Items} and in the online replication package. We evaluated every construct using uni-dimensional items, with responses captured on a 7-point Likert scale to denote the extent of agreement.

\subsubsection{Survey Data Collection}

To ensure the reliability and validity of our research, we started by determining the essential sample size through an \textit{a priori} power analysis using the G*Power tool~\cite{faul2009statistical}. Factoring in an effect size of 15\%, a significance level set at 5\%, and a desired power of 95\%, we established that a minimum of 204 respondents would be required, given the 16 predictors in our study.

For the data collection, we adopted a cluster sampling approach. This was carried out using the Prolific academic platform, renowned for its extensive user base of over 120,000 active participants. When juxtaposed with traditional methods like mailing lists, Prolific stands out for its heightened reliability, replicability, and data quality, making it especially favored within computer science research~\cite{russo2022recruiting}. 
Data collection took place in August 2021, coinciding with the conclusion of the semester. Participants were encouraged to complete the survey, offering insights based on their experiences during the Summer 2021 semester, in the middle of the COVID-19 lockdown.

The survey tool of choice was Qualtrics. To mitigate potential response biases, we employed a strategy of randomizing the sequence of questions within their respective blocks. Participants had to be at least 18 years old and to be actively enrolled in a Computer Science degree program.

We received a total of 311 responses. To enhance the reliability of our data, we incorporated three attention checks within the survey. Subsequently, 11 respondents who did not pass at least one of these checks were excluded from the dataset.

\subsubsection{Participant Demographics}

The cross-sectional study surveyed a total of 300 computer science students to evaluate their experiences with remote education during the COVID-19 lockdown. Participants were chosen from diverse backgrounds to ensure a comprehensive understanding of remote teaching practices from various perspectives.

\textbf{Country}: The majority of respondents hailed from the United States of America (38\%) and the United Kingdom (37.7\%). Other participants were from various countries, with India (2.7\%) and Nigeria (1.7\%) being the most represented among them.

\textbf{Gender Distribution}: Out of the total respondents, 55.7\% identified as men, 37.3\% as women, and a small percentage (0.7\%) identified as non-binary.

\textbf{Educational Background}: Most participants (74.3\%) were pursuing a Bachelor's degree. A smaller percentage were pursuing a Master's degree (12.7\%), while some identified their education level as ``Other" (5\%). A minority of the participants (1.7\%) were pursuing a PhD degree.

\textbf{Mode of Education}: A significant majority (76\%) of students mentioned that they attended their education always at home or remotely. The remaining 24\% indicated that they sometimes attended classes at home and sometimes at the university.

\textbf{Course Specialization}: The survey captured students from various computer science specializations. The most common courses among respondents were Programming Languages (21.3\%), Algorithms and Data Structures (13.3\%), and Software Engineering and Design (12.3\%).

\textbf{Digital Divide and Online Teamwork}: A noteworthy 79.7\% of students were familiar with the digital divide, indicating they had access to the necessary resources for online education. Moreover, 58.7\% of students collaborated with their peers through online teams during their remote education.

\textbf{Concluding Education Online}: Nearly half (49.7\%) of the participants wished their degree to be completed online, while the other half would prefer to conclude their studies on campus.\\

This diverse demographic profile allowed us to draw insights that are representative and applicable to a broader audience, ensuring the findings' generalizability across different backgrounds and experiences.

\section{Analysis}
\label{sec:analysis}

This section shows the results derived from our PLS-SEM analysis. We will first unpack the assessment of both the measurement and structural models, providing a comprehensive understanding of their interplay. Subsequently, we will explore the insights from the Multi-Group Analysis (MGA) and the Importance-Performance Map Analysis (IPMA), offering a holistic view of our findings and their implications.
To do so, we will follow the reporting guidelines by Russo \& Stol~\cite{russo2021pls}.

\subsection{Assessment of the Measurement Model}

To affirm the integrity and consistency of our framework, assessing the dependability of the latent variables is crucial. Thus, we initially explore the areas of discriminant validity, internal consistency reliability, and convergent validity.

\subsubsection{Discriminant Validity: Ensuring Construct Distinctiveness}

Discriminant validity measures the distinctiveness of one latent variable relative to another, helping determine if two constructs are genuinely distinct or if they encapsulate the same conceptual knowledge. For this assessment, we employed the Heterotrait-Monotrait ratio of correlations (HTMT), a method renowned for its superiority over alternatives such as the Fornell-Larcker criterion~\cite{henseler2015new}. Acceptable HTMT values should fall below 0.90~\cite{henseler2015new}.

Delving into our results, as illustrated in Table \ref{tab:HTMT1}, the majority of constructs exhibit clear demarcation. Yet, a closer inspection reveals an overlap between the constructs of `Fostering Relationships' and `Engagement', manifested in an HTMT value of 0.909. Additionally, `Organization' and `Expectation' show a similar intertwining with an HTMT value of 0.899. Both these pairs, exceed the preferred benchmark, suggesting that they might be conceptually overlapping.

Consequently, to refine our research model, we amalgamated these constructs. `Fostering-Relationships' and `Engagement' were consolidated into a singular construct named ``Engagement\_Fostering-Relationships", and similarly, `Organization' and `Expectation' were merged to form ``Organization\_Expectation".

\begin{table}[h]
\centering
\tiny
\robustify{\textbf}
\caption{Heterotrait-Monotrait ratio of correlations (HTMT) of the original model}
\label{tab:HTMT1}
\begin{tabular}{p{2cm} p{1cm} p{1cm} p{1cm} p{1cm} p{1cm} p{1cm} p{1cm} p{1cm} p{1cm}}
        \toprule
        & CO & EN & EX & FLEX & FR & LO & OR & TEC & TI \\
        \midrule
        Communication (CO) & & & & & & & & & \\
        Engagement (EN) & 0.824 & & & & & & & &  \\
        Expectation (EX) & 0.663 & 0.559 & & & & & & &  \\
        Flexibility (FLEX) & 0.801 & 0.843 & 0.760 & & & & & &  \\
        Fostering Relationships (FR) & 0.838 & \textbf{0.909} & 0.617 & 0.808 & & & & &  \\
        Learning Outcomes (LO) & 0.369 & 0.248 & 0.495 & 0.354 & 0.334 & & & &  \\
        Organization (OR) & 0.720 & 0.617 & \textbf{0.899} & 0.793 & 0.685 & 0.618 & & &  \\
        Technology (TEC) & 0.635 & 0.488 & 0.664 & 0.631 & 0.568 & 0.767 & 0.799 & &  \\
        Timeliness (TI) & 0.660 & 0.642 & 0.734 & 0.640 & 0.719 & 0.396 & 0.741 & 0.553 &  \\
        User Satisfaction (US) & 0.814 & 0.723 & 0.790 & 0.790 & 0.830 & 0.564 & 0.873 & 0.758 & 0.781  \\
        \bottomrule
\end{tabular}
\end{table}

Further validation is evident from Table \ref{tab:HTMT}, where all coefficients lie beneath the acceptable threshold. This underscores that each construct, in its revised form, maintains its distinct identity, symbolizing unique aspects of the study. This consolidation ensures a higher clarity and precision in capturing the nuances of the subject matter, bolstering the validity of our research findings.

\begin{table}[h]
\centering
\tiny
\robustify{\textbf}
\caption{Heterotrait-Monotrait ratio of correlations (HTMT) of the adapted model}
\label{tab:HTMT}
\begin{tabular}{p{5cm} p{1cm} p{1cm} p{1cm} p{1cm} p{1cm} p{1cm} p{1cm}}
        \toprule
        & CO & FLEX & FR-EN & LO & OR-EX & US & TEC \\
        \midrule
        Communication (CO) & & & & & & &  \\
        Flexibility (FLEX) & 0.801 & & & & & &  \\
        Fostering Relationship-Engagement (FR-EN) & 0.852 & 0.834 & & & & & \\
        Learning Outcomes (LO) & 0.369 & 0.354 & 0.329 & & & &  \\
        Organization-Expectation (OR-EX) & 0.718 & 0.788 & 0.677 & 0.566 & & &  \\
        Student Satisfaction (US) & 0.814 & 0.790 & 0.846 & 0.564 & 0.851 & &  \\
        Technology (TEC) & 0.635 & 0.631 & 0.580 & 0.767 & 0.719 & 0.758 &  \\
        Timeliness (TI) & 0.660 & 0.640 & 0.729 & 0.396 & 0.774 & 0.781 & 0.553 \\
        \bottomrule
    \end{tabular}
\end{table}

\subsubsection{Internal Consistency Reliability}

This assessment aims to validate that the latent variables are being measured with consistency and dependability. In our analysis, we look to the values of Cronbach's Alpha, Composite reliability rho\_a, and rho\_c, as depicted in Table \ref{tab:ICR}. These metrics should all surpass the threshold of 0.60, according to established standards~\cite{nunnally1978psychometric}. From our findings, it can be inferred that our evaluations satisfy the required reliability criteria.

\begin{table}[h]
\centering
\small
\caption{Internal consistency reliability}
\label{tab:ICR}
    \begin{tabular}{p{6cm} p{2cm} p{1cm} p{1cm} p{1cm} }
        \toprule
        & Cronbach's alpha & rho\_a & rho\_c & AVE \\
        \midrule
        Communication & 0.836 & 0.837 & 0.901 & 0.753 \\
        Flexibility & 0.831 & 0.831 & 0.899 & 0.747 \\
        Fostering Relationship-Engagement & 0.896 & 0.913 & 0.928 & 0.764 \\
        Learning Outcomes & 0.923 & 0.929 & 0.945 & 0.812 \\
        Organization-Expectation & 0.887 & 0.891 & 0.917 & 0.689 \\
        Student Satisfaction & 0.864 & 0.885 & 0.908 & 0.713 \\
        Technology & 0.722 & 0.748 & 0.841 & 0.638 \\
        Timeliness & 0.780 & 0.815 & 0.872 & 0.697 \\
        \bottomrule
    \end{tabular}
\end{table}

\subsubsection{Convergent Validity}

The final stage of validity evaluation focuses on determining the correlation levels between various elements and their associated constructs. it is important to emphasize that our hidden variables are measured reflectively (Mode A)\footnote{For an in-depth comparison of reflective and formative measurements, refer to Russo \& Stol~\cite{russo2021pls}.}. As a consequence, these signs should exhibit a significant proportion of variance by converging on their underlying variables. Two specific tests were utilized to confirm this notion.

The initial test centers on the average variance extracted (AVE), which needs to attain a value greater than 0.5, as stipulated by Hair et al.~\cite{hair2016pls}, as shown in Table \ref{tab:ICR}. 

The subsequent test ensures that the outer loadings for each measurement model related to the latent variable explain a minimum of 50\% of the variance. This evaluation is performed by gauging the reliability of the indicator, which must surpass the square root of 50\%, or 0.7. 
Table \ref{tab:OuterLoadings} consolidates the outcomes of the reliability of the indicators through cross-loadings. Any items failing to significantly contribute to the variance were omitted from our model during the analysis phase, with a full list of such items provided in Table \ref{tab:Items}. This exclusion led to an observable enhancement in the AVE, further strengthening the integrity of the model.

\begin{table}[ht!]
\centering
\small
\robustify{\textbf}
\caption{Cross loadings (full list of items in Table \ref{tab:Items})}
\label{tab:OuterLoadings}
\begin{tabular}{p{2cm} p{1cm} p{1cm} p{1cm} p{1cm} p{1cm} p{1cm} p{1cm} p{1cm}}
        \toprule
        & CO & FLEX & FR-EN & LO & OR-EX & US & TEC & TI \\
        \midrule
        CO\_1 & \textbf{0.870} & 0.645 & 0.689 & 0.248 & 0.554 & 0.610 & 0.403 & 0.485 \\
        CO\_2 & \textbf{0.846} & 0.562 & 0.557 & 0.295 & 0.565 & 0.599 & 0.442 & 0.468 \\
        CO\_3 & \textbf{0.888} & 0.528 & 0.692 & 0.311 & 0.502 & 0.624 & 0.425 & 0.476 \\
        EN\_1 & 0.694 & 0.703 & \textbf{0.893} & 0.263 & 0.573 & 0.713 & 0.437 & 0.585 \\
        EX\_1 & 0.484 & 0.546 & 0.472 & 0.358 & \textbf{0.842} & 0.603 & 0.463 & 0.558 \\
        EX\_2 & 0.476 & 0.514 & 0.471 & 0.402 & \textbf{0.836} & 0.554 & 0.430 & 0.475 \\
        EX\_3 & 0.539 & 0.660 & 0.565 & 0.439 & \textbf{0.862} & 0.680 & 0.513 & 0.578 \\
        FLEX\_1 & 0.603 & \textbf{0.852} & 0.673 & 0.231 & 0.559 & 0.591 & 0.427 & 0.466 \\
        FLEX\_2 & 0.586 & \textbf{0.882} & 0.631 & 0.292 & 0.603 & 0.575 & 0.416 & 0.454 \\
        FLEX\_3 & 0.538 & \textbf{0.859} & 0.586 & 0.293 & 0.608 & 0.588 & 0.407 & 0.436 \\
        FR\_1 & 0.492 & 0.454 & \textbf{0.760} & 0.215 & 0.408 & 0.504 & 0.298 & 0.451 \\
        FR\_2 & 0.705 & 0.697 & \textbf{0.915} & 0.321 & 0.578 & 0.690 & 0.433 & 0.557 \\
        FR\_3 & 0.686 & 0.657 & \textbf{0.918} & 0.266 & 0.565 & 0.723 & 0.449 & 0.566 \\
        LO\_1 & 0.364 & 0.363 & 0.362 & \textbf{0.914} & 0.590 & 0.565 & 0.624 & 0.398 \\
        LO\_2 & 0.305 & 0.309 & 0.305 & \textbf{0.911} & 0.497 & 0.484 & 0.624 & 0.298 \\
        LO\_3 & 0.242 & 0.230 & 0.183 & \textbf{0.877} & 0.360 & 0.362 & 0.545 & 0.237 \\
        LO\_4 & 0.260 & 0.216 & 0.236 & \textbf{0.902} & 0.413 & 0.440 & 0.575 & 0.292 \\
        OR\_2 & 0.472 & 0.467 & 0.456 & 0.470 & \textbf{0.793} & 0.594 & 0.466 & 0.499 \\
        OR\_3 & 0.591 & 0.625 & 0.566 & 0.487 & \textbf{0.818} & 0.702 & 0.533 & 0.561 \\
        TEC\_1 & 0.409 & 0.384 & 0.387 & 0.405 & 0.443 & 0.454 & \textbf{0.785} & 0.348 \\
        TEC\_2 & 0.367 & 0.355 & 0.359 & 0.722 & 0.501 & 0.494 & \textbf{0.840} & 0.349 \\
        TEC\_3 & 0.410 & 0.435 & 0.391 & 0.383 & 0.449 & 0.503 & \textbf{0.769} & 0.314 \\
        TI\_1 & 0.275 & 0.298 & 0.384 & 0.266 & 0.513 & 0.432 & 0.323 & \textbf{0.731} \\
        TI\_2 & 0.578 & 0.481 & 0.599 & 0.354 & 0.589 & 0.637 & 0.433 & \textbf{0.904} \\
        TI\_3 & 0.479 & 0.510 & 0.547 & 0.230 & 0.515 & 0.545 & 0.285 & \textbf{0.860} \\
        US\_1 & 0.720 & 0.693 & 0.795 & 0.360 & 0.679 & \textbf{0.880} & 0.490 & 0.622 \\
        US\_2 & 0.570 & 0.572 & 0.636 & 0.460 & 0.643 & \textbf{0.897} & 0.522 & 0.542 \\
        US\_3 & 0.416 & 0.444 & 0.499 & 0.253 & 0.463 & \textbf{0.716} & 0.369 & 0.451 \\
        US\_4 & 0.626 & 0.547 & 0.605 & 0.645 & 0.746 & \textbf{0.873} & 0.634 & 0.575 \\
        \bottomrule
\end{tabular}
\end{table}

\subsection{Evaluation of the Structural Model}

\begin{figure}[ht!]
\centerline{\includegraphics[width=1\linewidth]{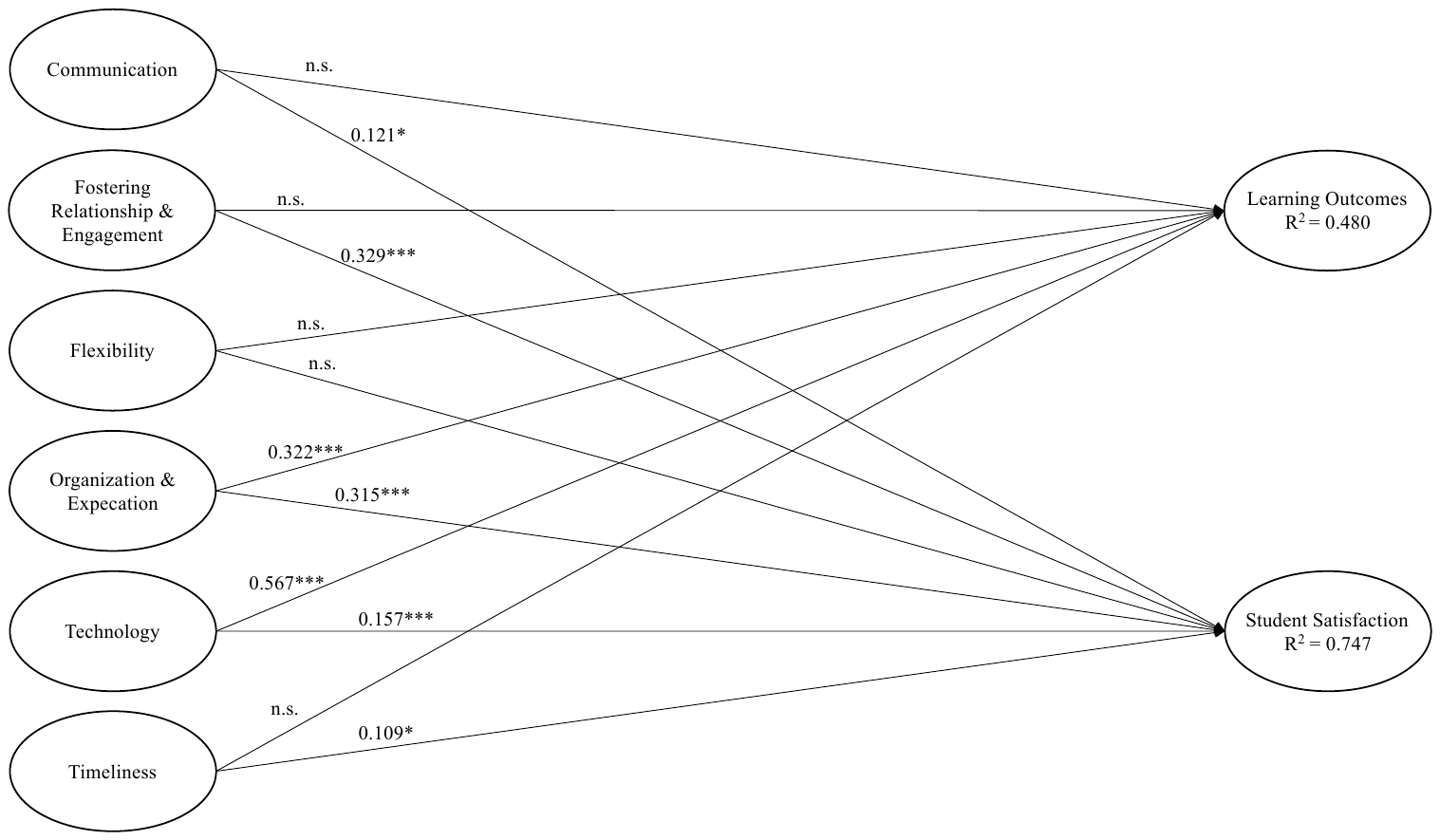}}
\caption{Structural model with $R^2$ and path coefficients (* \textit{p}$<0.05$, ** \textit{p}$<0.01$, *** \textit{p}$<0.001$, (NS) \textit{p}$>0.05$).}
\label{fig:StructuralModel}
\end{figure}

After ensuring the reliability of all our constructs through our Measurement Model assessment, we can now shift our focus towards evaluating the Structural Model, graphically represented in Figure~\ref{fig:StructuralModel}. This evaluation is pivotal in discussing the predictive power of our model and validating our research hypotheses.

\subsubsection{Collinearity Analysis}

We commenced our analysis by examining the correlation between the exogenous variables and the endogenous ones to ensure their independence, which is essential for preventing potential bias in our path estimates. The Variance Inflation Factor (VIF) is employed to detect multicollinearity, an extreme form of correlation. Ideally, the VIF values should remain below the threshold of five to avoid multicollinearity issues~\cite{Miles2014VIF}.

Our results present VIF values well within this range. For instance, the CO\_1 item registered a VIF of 2.046, while the US\_4 item had a VIF of 2.461. The highest value observed was for the LO\_1 item at 3.822, which is still comfortably below the threshold. These values indicate that our model is free from significant multicollinearity concerns.

\subsubsection{Path Relations: Significance and Relevance}

Path coefficients symbolize the theorized connections between latent variables. Being standardized, their values may vary from -1 to +1. Since PLS-SEM does not enforce distributional presumptions, parametric tests for assessing significance are not applicable. To circumvent this, a two-tailed bootstrapping method incorporating 10,000 subsamples with replacement is utilized.

The details of the bootstrapping results can be found in Table \ref{tab:PathCoeff}. This table provides an overview of the bootstrapping coefficients, mean, standard deviation, T statistics, and p-values corresponding to each relationship in our model.

Upon examination, our analysis uncovers that most of the relationships are statistically significant. Specifically, the relationships between Communication and Student Satisfaction, Fostering Relationship Engagement and Student Satisfaction, Organization Expectation and Learning Outcomes, Organization Expectation and Student Satisfaction, Technology and Learning Outcomes, and Technology and Student Satisfaction have \textit{p}-values less than 0.05, and their T statistics exceed 1.96. This is indicative of a 5\% significance level, affirming the relevance and importance of these connections in our model.

The results lend substantial support to our theoretical framework and reinforce the validity of the relationships between the constructs as initially hypothesized.

\begin{table}[h]
\centering
\small
\robustify{\textbf}
\caption{Path coefficients, bootstrap estimates, standard deviation, T statistics, and \textit{p}-values}
\label{tab:PathCoeff}
\begin{tabular}{p{4cm} p{2cm} p{2cm} p{1cm} p{1cm} p{1cm} }
        \toprule
Hypothesis & Coefficient  & Bootstrap Mean & St.Dev. & T & \textit{p}  \\
\midrule
CO \(\rightarrow\) LO & -0.038 & -0.040 & 0.080 & 0.476 & 0.634\\
CO \(\rightarrow\) US & 0.121 & 0.120 & 0.054 & 2.215 & 0.027\\
FLEX \(\rightarrow\) LO & -0.128 & -0.125 & 0.079 & 1.616 & 0.106\\
FLEX \(\rightarrow\) US & 0.009 & 0.009 & 0.055 & 0.158 & 0.875\\
FR-EN \(\rightarrow\) LO & -0.042 & -0.045 & 0.086 & 0.484 & 0.628\\
FR-EN \(\rightarrow\) US & 0.329 & 0.328 & 0.062 & 5.328 & 0.000\\
OR-EX \(\rightarrow\) LO & 0.322 & 0.317 & 0.083 & 3.874 & 0.000\\
OR-EX \(\rightarrow\) US & 0.315 & 0.318 & 0.060 & 5.240 & 0.000\\
TEC \(\rightarrow\) LO & 0.567 & 0.571 & 0.054 & 10.465 & 0.000\\
TEC \(\rightarrow\) US & 0.157 & 0.157 & 0.041 & 3.832 & 0.000\\
TI \(\rightarrow\) LO & 0.011 & 0.014 & 0.068 & 0.157 & 0.875\\
TI \(\rightarrow\) US & 0.109 & 0.109 & 0.048 & 2.250 & 0.024\\
\bottomrule
\end{tabular}
\end{table}

\subsubsection{Assessment of Determination Coefficients}

Upon successfully validating the statistical significance of most of our hypothesized relationships, we transition to the concluding part of our analysis, focusing on the predictive capabilities of the endogenous constructs. These capabilities are delineated in Table \ref{tab:R2}. The proportion of explained variance (\(R^2\)) in the dependent constructs is employed to measure the predictive strength. \(R^2\) illustrates the fraction of the variation in the dependent variable that is accountable by the independent variables. Since \(R^2\) tends to inflate with an increase in predictors, the adjusted \(R^2\) is also evaluated to account for the number of predictors in the model. These values fall within a 0 to 1 range. Although setting a universal threshold for \(R^2\) may be complicated, as it often depends on the specific subject matter~\cite{hair2016pls}, a common guideline suggests it should be above 0.19~\cite{chin1998PLS}. In our study, the values for Learning Outcomes and Student Satisfaction are 0.480 and 0.747, respectively, for \(R^2\), and 0.469 and 0.741 for the adjusted \(R^2\), surpassing the general guideline, thereby reflecting a robust predictive ability of the model, especially in regards to Student Satisfaction.

\begin{table}[h]
\centering
\small
\robustify{\textbf}
\caption{Coefficients of determination}
\label{tab:R2}
\begin{tabular}{p{5cm} p{4cm} p{4cm}}
        \toprule
Construct & \(R^2\) & \(R^2\) Adjusted \\
\midrule
Learning Outcomes & 0.480 & 0.469 \\
Student Satisfaction & 0.747 & 0.741 \\
\bottomrule
\end{tabular}
\end{table}

\subsubsection{Assessment of Predictive Performance}

In alignment with our research's main goal of making predictions rather than establishing causality, we executed a predictive evaluation using PLSpredict~\cite{shmueli2016elephant}. This method assesses whether a model, constructed with a training sample, can accurately predict the outcomes in a test sample. We divided our sample into ten segments and conducted ten repetitions to obtain the PLSpredict statistics. The interpretation of the results followed the principles laid out by Shmueli et al.~\cite{shmueli2019predictive}, as illustrated in Table \ref{tab:PLSpredict}. Remarkably, all latent variables showcased a highly positive \textit{Q}\textsuperscript{2}\(_{\text{predict}}\), signifying the model's substantial predictive capability. For instance, Learning Outcomes and Student Satisfaction registered \textit{Q}\textsuperscript{2}\(_{\text{predict}}\) values of 0.448 and 0.729, respectively, which further underscores the robustness of our predictive model.

\begin{table}[h]
\centering
\small
\robustify{\textbf}
\caption{Constructs prediction summary}
\label{tab:PLSpredict}
\begin{tabular}{p{6cm} p{2cm} p{2cm} p{2cm}}
        \toprule
Construct & \textit{Q}\textsuperscript{2}\(_{\text{predict}}\) & RMSE & MAE \\
\midrule
Learning Outcomes & 0.448 & 0.748 & 0.612 \\
Student Satisfaction & 0.729 & 0.524 & 0.391 \\
\bottomrule
\end{tabular}
\end{table}

\subsubsection{Examining Predictive Influence}

In this final phase of our analysis, the focus is on the assessment of the predictive influence of our model by evaluating the effect sizes, denoted as \( f^2 \), as detailed in Table \ref{tab:f2}. Effect sizes measure the strength of various relationships within the model, with threshold values set at 0.02, 0.15, and 0.35 for small, medium, and large effects, respectively~\cite{Cohen1988PowerAnalysis}.

The table reveals that the relationship between Technology and Learning Outcomes has the most substantial effect, with an \( f^2 \) value of 0.389, which falls into the category of a large effect. Other constructs exhibit either small or negligible effects on Learning Outcomes and Student Satisfaction. Notably, Organization-Expectation and Fostering Relationship-Engagement also demonstrate some influence on Student Satisfaction, with medium effect sizes. Thus, the analysis underscores the role of Technology in predicting Learning Outcomes, while other factors may have a more nuanced impact on the dependent variables.

\begin{table}[h]
\centering
\small
\robustify{\textbf}
\caption{Effect sizes ($f^2$)}
\label{tab:f2}
\begin{tabular}{p{6cm} p{3cm} p{3cm}}
        \toprule
Constructs & Learning Outcomes & Student Satisfaction \\
\midrule
Communication               & 0.001 & 0.022 \\
Flexibility                 & 0.012 & 0.000 \\
Fostering Relationship-Engagement & 0.001 & 0.135 \\
Organization-Expectation    & 0.073 & 0.143 \\
Technology                  & 0.389 & 0.061 \\
Timeliness                  & 0.000 & 0.023 \\
\bottomrule
\end{tabular}
\end{table}

\subsection{Assessing Influential Constructs for Student Satisfaction and Learning Outcomes in Remote Teaching: An Exploration using Importance-Performance Map Analysis}

In the unique context of the COVID-19 pandemic and the resulting shift to remote education, the assessment of Student Satisfaction and Learning Outcomes has taken on unprecedented significance. This paper employs the Importance-Performance Map Analysis (IPMA) methodology to explore the contributing factors, offering actionable insights for educators, administrators, and policymakers.

The following subsubsections provide an in-depth analysis of the constructs affecting Student Satisfaction and Learning Outcomes, presenting a comprehensive view informed by the PLS-SEM investigation.

\subsubsection{Uncovering Key Drivers for Student Satisfaction: An Examination through Importance-Performance Map Analysis}

Our analysis begins with an exploration of the factors influencing Student Satisfaction in remote teaching, using the IPMA approach. Table \ref{tab:LV_Performance} reveals the performance of various constructs, including Communication, Flexibility, Fostering Relationship-Engagement, Organization-Expectation, Technology, and Timeliness, all of which surpass the 62\% mark.

Examining the importance of individual constructs (Table \ref{tab:UnTotalEffects}), Fostering Relationship-Engagement and Organization-Expectation stand out as the most influential, emphasizing their role in enhancing student satisfaction in the remote education landscape.

\begin{figure}[!ht]
\centering
\includegraphics[width=0.8\linewidth]{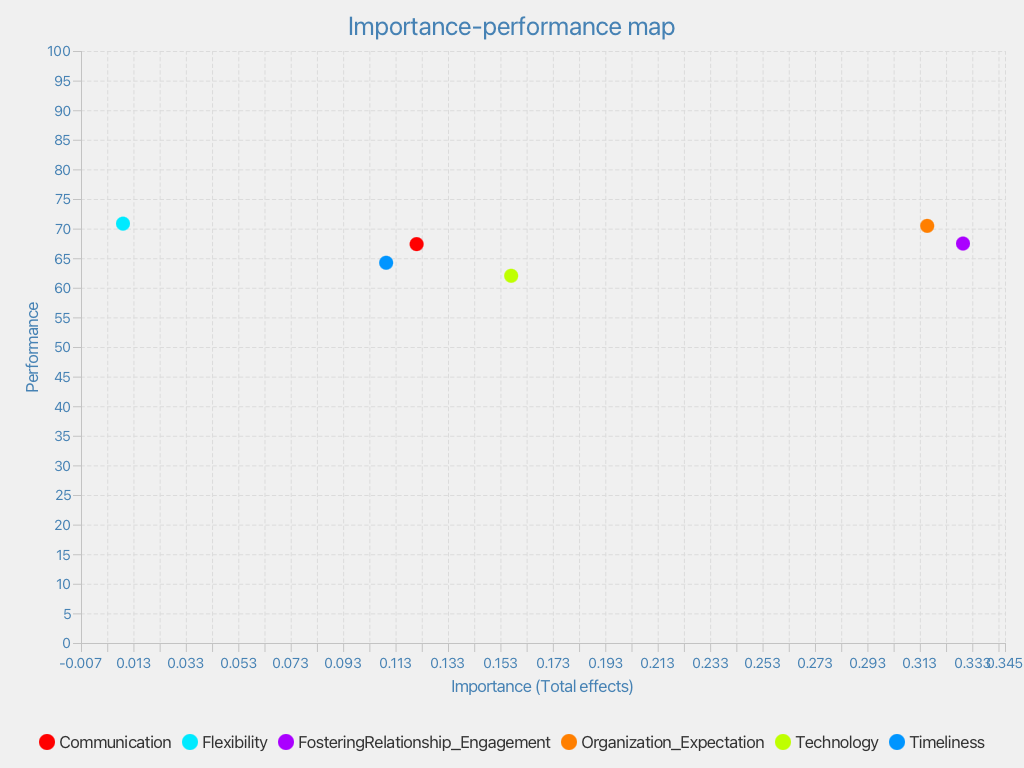}
\caption{Importance-Performance Map Analysis of Student Satisfaction.}
\label{fig:IPMA-US}
\end{figure}

Figure \ref{fig:IPMA-US} maps the interplay between the importance and performance, providing strategic insights. It implies that efforts to boost Student Satisfaction may benefit from focusing on relationship engagement and organizational expectation. Conversely, Flexibility and Timeliness might be of lesser priority in this context.

\begin{table}[!ht]
\centering
\sisetup{
    group-digits=true,
    group-minimum-digits=4,
    table-format=0.3,
    mode=text,
    detect-weight=true, 
    detect-family=true
}
\small
\robustify{\textbf}
\caption{Constructs Performance for Student Satisfaction}
\label{tab:LV_Performance}
\begin{tabular}{p{6cm} p{4cm}}
        \toprule
Construct & Construct Performances        \\
    \midrule
Communication & 67.378 \\ 
Flexibility & 70.840 \\
Fostering Relationship-Engagement & 67.477 \\
Organization-Expectation & 70.462 \\
Technology & 62.032 \\
Timeliness & 64.231 \\
\bottomrule
\end{tabular}
\end{table}

\begin{table}[!ht]
\centering
\sisetup{
    group-digits=true,
    group-minimum-digits=4,
    table-format=0.3,
    mode=text,
    detect-weight=true, 
    detect-family=true
}
\small
\robustify{\textbf}
\caption{Constructs Importance (Unstandardized Total Effects) for Student Satisfaction}
\label{tab:UnTotalEffects}
\begin{tabular}{p{6cm} p{4cm}}
        \toprule
Construct & Student Satisfaction \\
    \midrule
Communication & 0.121 \\ 
Flexibility & 0.009 \\
Fostering Relationship-Engagement & 0.329 \\
Organization-Expectation & 0.315 \\
Technology & 0.157 \\
Timeliness & 0.109 \\
\bottomrule
\end{tabular}
\end{table}

\subsubsection{Analyzing Key Influencers for Learning Outcomes: A Study using Importance-Performance Map Analysis}

This segment of the investigation delves into the factors affecting Learning Outcomes, using the IPMA methodology. Table \ref{tab:LO_Performance} reveals strong performance across all identified constructs, with Technology emerging as the most influential factor.

The importance of individual constructs, as displayed in Table \ref{tab:LO_TotalEffects}, ranges from -0.128 to 0.567, highlighting the role of technological integration in remote teaching and its significant influence on Learning Outcomes.

\begin{figure}[!ht]
\centering
\includegraphics[width=0.8\linewidth]{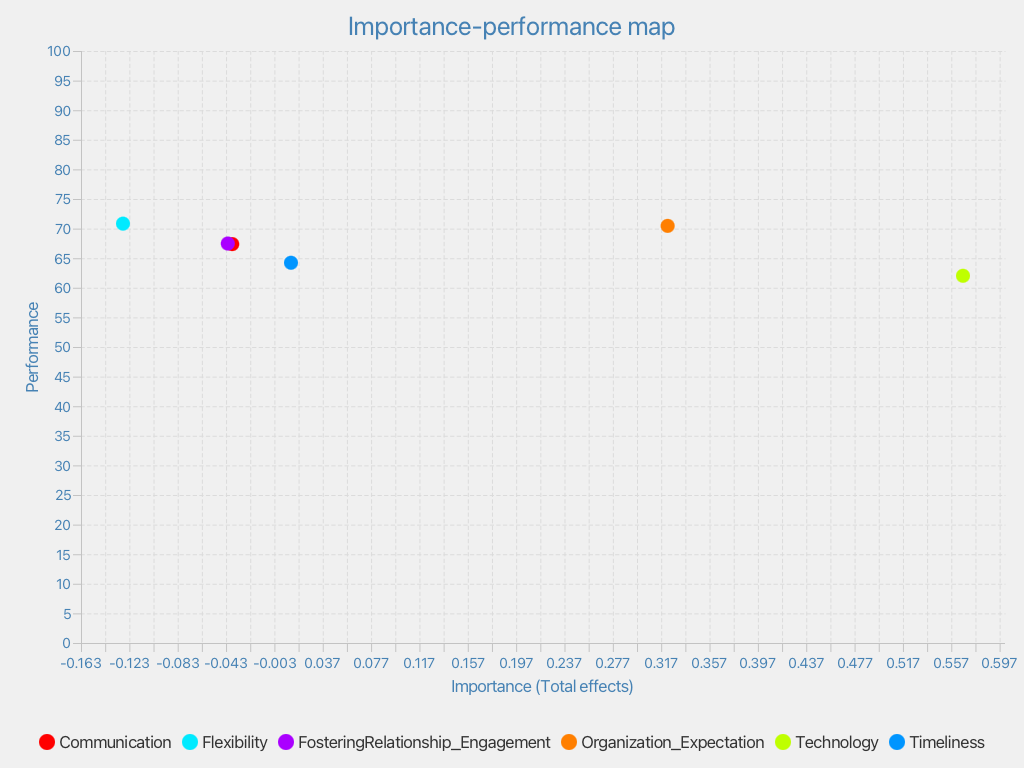}
\caption{Importance-Performance Map Analysis of Learning Outcomes.}
\label{fig:IPMA-LO}
\end{figure}

A conceptual diagram (similar to Figure \ref{fig:IPMA-LO}) could further elucidate the interaction between the importance and performance, suggesting that emphasis on amplifying Technology could lead to improved Learning Outcomes, while other constructs might appear less pivotal.
\\

In conclusion, our nuanced IPMA analysis presents a comprehensive and data-driven understanding of remote teaching practices during the COVID-19 pandemic. By  examining the complex interplay between various factors such as Communication, Flexibility, Fostering Relationship-Engagement, Organization-Expectation, Technology, and Timeliness, we have unveiled their collective influence on Student Satisfaction and Learning Outcomes.
The insights derived from this analysis not only reflect the performance of these constructs but also highlight the areas that are most significant in contributing to Student Satisfaction. For instance, the emphasis on Fostering Relationship-Engagement and Organization-Expectation has proven essential in enhancing student contentment in a remote learning environment.
In the context of Learning Outcomes, our analysis accentuates the critical role of Technology, suggesting that the thoughtful integration of technological tools and platforms can substantially impact the effectiveness of online education. The negative values associated with some constructs also provide a cautionary note, indicating areas that might hinder the attainment of desired Learning Outcomes.

\begin{table}[!ht]
\centering
\sisetup{
    group-digits=true,
    group-minimum-digits=4,
    table-format=0.3,
    mode=text,
    detect-weight=true,
    detect-family=true
}
\small
\robustify{\textbf}
\caption{Constructs Performance for Learning Outcomes}
\label{tab:LO_Performance}
\begin{tabular}{p{6cm} p{4cm}}
        \toprule
Construct & Construct Performances        \\
    \midrule
Communication & 67.378 \\
Flexibility & 70.840 \\
Fostering Relationship-Engagement & 67.477 \\
Organization-Expectation & 70.462 \\
Technology & 62.032 \\
Timeliness & 64.231 \\
\bottomrule
\end{tabular}
\end{table}

\begin{table}[!ht]
\centering
\sisetup{
    group-digits=true,
    group-minimum-digits=4,
    table-format=0.3,
    mode=text,
    detect-weight=true,
    detect-family=true
}
\small
\robustify{\textbf}
\caption{Constructs Importance (Unstandardized Total Effects) for Learning Outcomes}
\label{tab:LO_TotalEffects}
\begin{tabular}{p{6cm} p{4cm}}
        \toprule
Construct & Learning Outcomes \\
    \midrule
Communication & -0.038 \\
Flexibility & -0.128 \\
Fostering Relationship-Engagement & -0.042 \\
Organization-Expectation & 0.322 \\
Technology & 0.567 \\
Timeliness & 0.011 \\
\bottomrule
\end{tabular}
\end{table}

\subsection{Delineating Group-Specific Variations: A Multi-Group Analysis}

The abrupt change to remote teaching during the COVID-19 pandemic presented a unique and uncharted challenge for educators and students alike. While our primary analysis painted a comprehensive landscape of this shift, delving deeper into specific clusters could provide more nuanced insights. Recognizing this, we employed the Multi-Group Analysis (MGA) to ascertain if specific subgroups had distinct experiences or shared a common narrative.

MGA, with its bootstrapping-centric non-parametric methodology, excels in isolating group-specific outcomes, offering a finer lens to discern variations~\cite{henseler2009use,sarstedt2011multigroup}. In this exploration, we concentrated on four pivotal dimensions:

\begin{itemize}
    \item \textbf{Geographical Location}: Comparing experiences between students from the US and the UK, this analysis aimed to understand if regional distinctions influenced remote learning perceptions. We did so because most of our informants are located in one of these countries.
    \item \textbf{Academic Stage}: By contrasting the feedback from Bachelor's and Master's students, we aimed to discern any variations arising from curriculum complexities or academic expectations.
    \item \textbf{Gender}: This dimension was pivotal in uncovering any gender-specific patterns or disparities in the remote learning experience, informing institutions about inclusivity aspects. Unfortunately, for statistical significance, we could only include man or women subjects in this analysis. 
    \item \textbf{Team Dynamics}: Given the collaborative nature of many academic projects, we evaluated if students who worked in teams navigated remote learning differently from those accustomed to individual tasks.
\end{itemize}

The insights from the MGA were as follows: there were no significant differences across all examined dimensions. This homogeneity in experiences, irrespective of geography, academic stage, gender, or collaboration preferences, suggests a shared journey through the challenges and opportunities of remote education during the pandemic.

While this uniformity might be indicative of the global nature of the challenges posed by remote teaching—technological hurdles, adaptation to virtual tools, and the isolation of digital learning—it also emphasizes a collective resilience and adaptability among students. This shared experience, as revealed by the feedback from Bachelor's and Master's students, underscores the universal nature of these challenges. The alignment of perceptions across gender lines and team dynamics further accentuates the all-encompassing nature of the remote teaching experience during the pandemic.
Yet, as we interpret these collective findings, it is paramount to acknowledge that within these overarching narratives, individual subgroups might challange unique concerns and experiences. 

A thorough breakdown of our MGA findings, along with detailed analytical procedures, is available in the Online Supplementary Materials. Interested readers can access this comprehensive resource via the provided replication package link\footnote{Access the replication package here: https://doi.org/10.5281/zenodo.8268924.}.

\subsubsection{Digital Divide: The Critical Influence of Internet Quality on Remote Education}

The COVID-19 pandemic's transition to remote teaching underscored the pivotal role of reliable internet connectivity in the online education landscape. While various factors influence students' remote learning experiences, the digital divide – encapsulating disparities in digital resources access and quality – stands out as a potential determinant. Recognizing this, we employed Multi-Group Analysis (MGA) to compare students' experiences based on their internet quality, differentiating between those with optimal connectivity and those with inadequate access, as in Table \ref{tab:MGA}.

Of all the MGA analyses we conducted in our study, this specific exploration into the digital divide yielded the only significant results. Relationships such as `Fostering Relationship-Engagement \(\rightarrow\) Learning Outcomes', and `Technology \(\rightarrow\) Student Satisfaction' display significant variations between the two groups. This indicates that internet quality significantly influences these aspects of students' remote learning experiences.

Placing this into context, the sudden shift to remote learning meant that students' reliance on internet connectivity dramatically surged. Activities such as attending virtual classes, downloading study materials, and collaborative group work hinged on the quality of internet access. The fact that students with good internet connectivity reported different levels of satisfaction in communication and technology underscores the vital role of uninterrupted, quality internet in fostering optimal remote learning experiences.

However, it is essential to highlight that while the digital divide played a role in certain aspects of the learning experience, for many constructs, the differences were not statistically significant. This suggests that students, irrespective of their internet quality, faced shared challenges and benefitted from similar teaching practices during remote education.

In summary, while various factors influenced remote learning experiences during the pandemic, the digital divide emerged as a significant determinant in specific areas like communication and technology use. These findings stress the importance of ensuring equitable digital access and adapting teaching methods to cater to students across varying levels of internet connectivity. As educational institutions strategize for future remote or hybrid teaching models, addressing this digital divide will be paramount to ensure an inclusive, effective, and satisfactory learning experience for all students.

\begin{table}[h]
\centering
\sisetup{
    group-digits=true,
    group-minimum-digits=4,
    table-format=0.3,
    mode=text,
    detect-weight=true, 
    detect-family=true
}
\small
\robustify{\textbf}
\caption{Multi-Group Analysis: Digital Divide}
\label{tab:MGA}
\begin{tabular}{lp{2.5cm}p{1.5cm}}
        \toprule
Hypothesis & Path Coeff. diff  &  \textit{p}-value  \\
    \midrule
Communication \(\rightarrow\) Learning Outcomes & 0.0965 & 0.3799 \\
Communication \(\rightarrow\) Student Satisfaction & 0.1833 & 0.086 \\
Flexibility \(\rightarrow\) Learning Outcomes & 0.073 & 0.4965 \\
Flexibility \(\rightarrow\) Student Satisfaction & -0.148 & 0.2604 \\
Fostering Relationship-Engagement \(\rightarrow\) Learning Outcomes & -0.482 & \textbf{0.044} \\
Fostering Relationship-Engagement \(\rightarrow\) Student Satisfaction & -0.198 & 0.1431 \\
Organization-Expectation \(\rightarrow\) Learning Outcomes & 0.059 & 0.5806 \\
Organization-Expectation \(\rightarrow\) Student Satisfaction & 0.2403 & 0.051 \\
Technology \(\rightarrow\) Learning Outcomes & 0.062 & 0.5201 \\
Technology \(\rightarrow\) Student Satisfaction & -0.243 & \textbf{0.028} \\
Timeliness \(\rightarrow\) Learning Outcomes & 0.016 & 0.6132 \\
Timeliness \(\rightarrow\) Student Satisfaction & -0.075 & 0.3979 \\
\bottomrule
\end{tabular}
\end{table}

\section{Discussion}
\label{sec:discussion}

The unprecedented shift to online education during the COVID-19 pandemic has brought forth a myriad of challenges and opportunities for educators and students alike. This study, rooted in the experiences of computer science students exposed to a year of remote education, sought to provide actionable insights into effective remote teaching practices. The findings, as summarized in Table \ref{tab:Implications}, reveal that while remote teaching practices may not significantly influence learning outcomes (with an $R^2$ of 48\%), they play a crucial role in enhancing student satisfaction (with an impressive $R^2$ of 75\%).

\begin{table}[!ht]
\centering
\small
\robustify{\textbf}
\caption{Summary of findings and implications}
\label{tab:Implications}
\begin{tabular}{p{3cm}p{3cm}p{6cm}}
        \toprule
Hypothesis & Findings  &  Implications   \\
    \midrule
\textit{H1}: Communication \(\rightarrow\) Learning Outcomes & Not supported. &  While effective communication is foundational in online teaching, it may not be the only factor influencing learning outcomes. A holistic approach encompassing other pedagogical elements might be necessary. \\
   \addlinespace
\textit{H2}: Communication \(\rightarrow\) Students Satisfaction  & Supported. Path coefficient: 0.121; Effect size: 0.022. & Emphasizing clear and empathetic communication can significantly boost student satisfaction, fostering a more conducive online learning environment. \\
   \addlinespace
\textit{H3}: Flexibility \(\rightarrow\) Learning Outcomes  & Not supported. &  Flexibility, while a valuable trait in online teaching, might not be directly correlated with perceived learning outcomes. Its impact could be more indirect or combined with other factors. \\
   \addlinespace
\textit{H4}: Flexibility \(\rightarrow\) Students Satisfaction  & Not supported. & While flexibility is appreciated, it might not be the primary driver of student satisfaction. Other elements, such as communication or organization, might have a more direct impact. \\
   \addlinespace
\textit{H5}: Fostering Relationships-Engagement \(\rightarrow\) Learning Outcomes & Not supported. & Even though fostering relationships is essential, it might not have a direct correlation with learning outcomes. However, its indirect effects, such as creating a supportive learning environment, are undeniable. \\
   \addlinespace
\textit{H6}: Fostering Relationships-Engagement \(\rightarrow\) Students Satisfaction  & Supported. Path coefficient: 0.329; Effect size: 0.135. & Cultivating strong relationships and promoting engagement can lead to a more satisfied and engaged student body in online courses. \\
   \addlinespace
\textit{H7}: Organization-Expectation \(\rightarrow\) Learning Outcomes  & Supported. Path coefficient: 0.322; Effect size: 0.073. IPMA: Highest performance for Learning Outcomes. & A meticulously organized online course, combined with clear expectations, can significantly enhance the learning experience. The IPMA results underscore the paramount importance of organization in boosting learning outcomes. \\
   \addlinespace
\textit{H8}: Organization-Expectation \(\rightarrow\) Students Satisfaction  & Supported. Path coefficient: 0.315; Effect size: 0.143. & A well-structured course with clear expectations can significantly elevate student satisfaction, as it provides clarity and reduces potential frustrations. \\
   \addlinespace
\textit{H9}: Technology \(\rightarrow\) Learning Outcomes   & Supported. Path coefficient: 0.567; Effect size: 0.389. & Leveraging technology effectively can transform the online learning experience, making content more accessible, engaging, and interactive, leading to enhanced learning outcomes. \\
   \addlinespace
\textit{H10}: Technology \(\rightarrow\) Students Satisfaction   & Supported. Path coefficient: 0.157; Effect size: 0.061. IPMA: Most important for Students Satisfaction. & The adept integration of technology is pivotal for student satisfaction. As per the IPMA results, technology stands out as the most influential factor in determining student satisfaction in online courses. \\
   \addlinespace
\textit{H11}: Timeliness \(\rightarrow\) Learning Outcomes  & Not supported. & Timely feedback and interactions, while essential, might not be the direct influencers of learning outcomes. However, they play a crucial role in maintaining engagement and momentum in the course. \\
   \addlinespace
\textit{H12}: Timeliness \(\rightarrow\) Students Satisfaction  & Supported. Path coefficient: 0.109; Effect size: 0.023. & Swift responses and timely feedback can instill confidence in students, making them feel valued and enhancing their overall satisfaction in the course. \\
\bottomrule
\end{tabular}
\end{table}

\subsection{Factors Influencing Student Satisfaction and Learning Outcomes}

\subsubsection{Communication}

\paragraph{Impact on Student Satisfaction}
During the pandemic, effective communication emerged as a significant factor in boosting student satisfaction in online courses. The absence of physical interactions accentuated the importance of clear and empathetic communication in fostering a conducive online learning environment. To address this, educators should prioritize establishing clear communication channels. Regular virtual check-ins, dedicated Q\&A sessions, and feedback mechanisms can ensure students feel connected and informed.

\paragraph{Impact on Learning Outcomes}
While communication was foundational in the remote teaching necessitated by the pandemic, it may not have been the sole determinant of learning outcomes. To enhance learning outcomes, educators could consider blending synchronous and asynchronous communication tools and promoting collaborative learning through group discussions and projects.

\subsubsection{Re-evaluating the Role of Flexibility}
While flexibility in online education is often lauded as a significant advantage, our findings present a nuanced view. Although students appreciate the flexibility inherent in online courses, it does not emerge as a primary driver of satisfaction or perceived learning efficacy. This observation contrasts with the assertions of Ke \& Kwak~\cite{ke2013}, who emphasize the role of flexibility in enhancing student satisfaction. Our results suggest that while flexibility is valued, other factors, such as engagement and clear communication, may hold more weight in determining overall satisfaction and learning outcomes.

\subsubsection{Fostering Relationship-Engagement}

\paragraph{Impact on Student Satisfaction}
Building a sense of community became even more crucial during the pandemic. Organizing virtual community-building activities, peer mentorship programs, and fostering open discussions can help in building strong relationships and promoting engagement.

\paragraph{Impact on Learning Outcomes}
Despite the emphasis on fostering relationships during the pandemic, it might not have had a direct relation with learning outcomes. 

\subsubsection{Organization-Expectation}

\paragraph{Impact on Student Satisfaction}
The pandemic highlighted the importance of a well-structured online course with clear expectations. To address this, educators could provide a detailed and clear syllabus at the start, ensure consistency in course layout and materials, and set clear milestones for students.

\paragraph{Impact on Learning Outcomes}
The findings from this period underscore the importance of organization in boosting learning outcomes. Implementing progress trackers and providing students with clear guidelines on what is expected of them can optimize learning outcomes.

\subsubsection{Technology}

\paragraph{Impact on Student Satisfaction}
The adept integration of technology became even more pivotal during the pandemic. Ensuring that the learning management system is intuitive, offering robust tech support, and integrating interactive tools can significantly elevate the online learning experience.

\paragraph{Impact on Learning Outcomes}
The pandemic emphasized the role of technology in transforming the online learning experience. Incorporating interactive tools like virtual labs and offering content in diverse formats can cater to different learning styles, further enhancing the learning experience.

\subsubsection{Timeliness}

\paragraph{Impact on Student Satisfaction}
Timely feedback and responses were essential during the pandemic. Educators could ensure timely grading and feedback and schedule regular office hours for real-time student interactions.

\paragraph{Impact on Learning Outcomes}
While timeliness was essential during the pandemic, it might not have been the direct influencer of learning outcomes.

\subsection{Delving Deeper: Insights from Importance-Performance Map Analysis (IPMA)}

The Importance-Performance Map Analysis (IPMA) offers a nuanced lens through which we can understand the dynamics of online learning. By juxtaposing the importance and performance of various constructs, we can derive actionable insights that can inform pedagogical strategies in online education.

\paragraph{The Pivotal Role of Technology}
A standout observation from our IPMA is the pronounced influence of Technology on both student satisfaction and learning outcomes. In the digital age, the efficacy of online education is inextricably linked to the technological tools employed. Our findings underscore the significance of integrating robust and user-friendly technological platforms to facilitate learning. This observation aligns with the work of Anderson \& Dron~\cite{anderson2011}, who emphasize the transformative potential of technology in shaping online educational experiences. For disciplines like computer science, the importance of technology is further magnified. Practical exercises, simulations, and real-time feedback are integral components of computer science education, necessitating the deployment of advanced technological tools.

\paragraph{Organization-Expectation: A Key to Streamlined Learning}
Another salient insight from our IPMA is the role of Organization-Expectation in determining learning outcomes. A coherent and well-structured online course acts as a beacon, guiding students through the learning journey. Our findings resonate with the research of Margaryan et al.~\cite{margaryan2015}, who posit that clear instructional design and course organization are pivotal in enhancing student comprehension and engagement. For educators, this underscores the need to meticulously plan and structure online courses. A transparent course roadmap, complete with learning objectives, weekly modules, and assessment criteria, can significantly elevate the online learning experience, making content more digestible and the learning trajectory more predictable.

\subsection{Unraveling the Tapestry of Student Experiences: Insights from Multi-Group Analysis (MGA)}

The Multi-Group Analysis (MGA) serves as a powerful tool to understand the heterogeneity or homogeneity of experiences across different student subgroups. By segmenting our sample based on various criteria, we aimed to uncover nuanced insights that might be obscured in a more aggregated analysis.

\paragraph{A Collective Journey through Digital Pedagogy}
Our Multigroup Analysis (MGA) unveils a compelling uniformity in the pedagogical experiences of students, irrespective of their geographical locale, academic progression, gender, or collaborative dynamics. This homogeneity in experiences underscores the universality of the challenges and successes encountered during the pandemic-induced transition to remote instruction. Such a pervasive consistency intimates that the exigencies of digital education during this period were not merely localized phenomena but rather global academic challenges that transcended traditional educational and personal demarcations.

\paragraph{The Digital Divide: Bridging the Gap}
The transition to online learning brought to the forefront a myriad of shared experiences among students. However, the digital divide emerged as a significant outlier in this collective journey. Differences in internet quality across regions and households highlighted the pivotal role of a stable and fast connection in the realm of online education. For many students, having a reliable internet connection was synonymous with active participation, seamless video streaming, and uninterrupted interactions with peers and instructors. Conversely, those with inconsistent or slow connections often grappled with video lags, dropped calls, and missed content, leading to feelings of isolation, frustration, and a sense of being left behind. This divide underscores the need for educational institutions and policymakers to prioritize digital infrastructure, ensuring that all students, regardless of their location or economic background, have equal access to quality online education.

\subsection{Implications for Educators}

The findings from our study, while rooted in the specific context of the COVID-19 pandemic's shift to remote education, offer several enduring lessons for educators. These insights can guide pedagogical strategies, not just in times of crisis, but also in the evolving landscape of education that increasingly embraces digital modalities.

\paragraph{Embracing Technology as a Pedagogical Ally}
Our study underscores the pivotal role of technology in shaping student satisfaction and learning outcomes. Educators should view technology not merely as a medium of instruction but as a pedagogical ally. Leveraging digital tools, from interactive platforms to simulation software, can make learning more engaging and effective. Continuous professional development in technology integration can empower educators to harness its full potential.

\paragraph{Prioritizing Clear Communication and Relationship Building}
The significance of clear communication and fostering relationships in our findings resonates with the foundational principles of education. In a virtual environment, these elements become even more crucial. Regular check-ins, feedback sessions, and open channels of communication can bridge the virtual gap. Building a sense of community in online classes, perhaps through group projects or discussion forums, can mitigate feelings of isolation.

\paragraph{Designing Structured and Organized Online Courses}
The importance of organization and clear expectations cannot be overstated. A well-structured online course, complete with a clear roadmap, learning objectives, and timely assessments, can significantly enhance the student learning experience. Tools like learning management systems can aid in organizing course content and tracking student progress.

\paragraph{Addressing the Digital Divide}
Our findings on the digital divide highlight a pressing concern. While educators might have limited influence over infrastructural issues, they can adopt inclusive teaching strategies. For instance, providing downloadable resources, asynchronous learning options, or even offline assignments can ensure that students with limited internet access are not left behind.

\paragraph{Continuous Feedback and Adaptation}
The dynamic nature of online education, coupled with the diverse needs of students, calls for a feedback-driven approach. Regular surveys, feedback forms, or even informal discussions can provide insights into what's working and what's not. Educators could be open to adapting their strategies based on this feedback, ensuring a responsive and student-centric approach.\\

In conclusion, the transition to remote education during the pandemic has illuminated several facets of online teaching and learning. While challenges abound, the opportunities for innovation and improvement are immense. By internalizing the lessons from this period and integrating them into their pedagogical toolkit, educators can be better prepared for the future of education, whatever it may hold.

\subsection{Threats to Validity}

To ensure the credibility and robustness of our research, we thoroughly evaluated potential threats to validity. To do so, we adopted a comprehensive framework grounded in the work of Wohlin et al.~\cite{Wohlin}.

\textbf{Internal Validity}.
We performed a cluster-randomized probability sampling strategy for validating our research model~\cite{gravetter2018research}. While surveying the global student body was logistically challenging, we focused on a subset, primarily sourcing participants from the Prolific platform. Although cluster sampling might be viewed as less granular than pure random sampling, it offered a pragmatic and cost-effective compromise. Our approach aligns with Baltes and Ralph's observations, highlighting the underuse of probability sampling in seminal software engineering research~\cite{baltes2020sampling}. Our sample comprised 300 computer science students. However, given the geographic concentration of our respondents (mainly located in the US or UK), our sample might not be universally representative of global CS students.

\textbf{External Validity}.
Our PLS-SEM analysis was oriented by the ambition to generalize our findings across diverse contexts~\cite{stol2018abc}. The robust sample size of 300 respondents, informed by a preceding power analysis, endowed our study with statistical strength. Additionally, the Multi-Group Analysis, which dissected responses across varied demographics, revealed minimal significant disparities, underscoring the universality of our findings.

\textbf{Construct Validity}.
Construct measurements were anchored in the perspectives of computer science students and were garnered through a singular informant approach. While our data was self-reported, and inherently subjective, we solicited participants' alignment with established scholarly statements. Recognizing the potential pitfalls of self-reporting, we integrated three attention checks into our survey, which flagged eleven responses. Our survey instrument, grounded in prior works, was randomized within blocks, thereby minimizing biases.

\textbf{Statistical Conclusion Validity}.
Our analytical arsenal was powered by SmartPLS (4.0.9.5), a statistical tool vouched for in over a thousand academic articles~\cite{ringle2015smartpls}. We diligently adhered to cutting-edge PLS-SEM best practices, ensuring that our statistical procedures stood up to the most recent guidelines~\cite{russo2021pls}.

\section{Conclusions}
\label{sec:Conclusions}

As we reflect upon the unprecedented challenges posed by the COVID-19 pandemic, it becomes evident that the world of education underwent a transformative shift. Universities, educators, and students were thrust into a sudden transition, adapting to a remote mode of instruction. This retrospective study aimed to shed light on the effectiveness of remote teaching practices during this period, providing insights that could guide future educational strategies.

Our findings underscored several key themes. The importance of technology as a cornerstone of effective online education was evident, emphasizing the need for its thoughtful integration into pedagogical practices. Clear communication and fostering relationships emerged as pivotal, highlighting the human element in a digital learning environment. The study also revealed the universal challenges faced by students, irrespective of geographical location, academic stage, or gender, pointing to a shared journey through the pandemic's educational landscape.

However, the digital divide, representing disparities in access to digital resources, emerged as a significant concern. As we move forward, it is imperative that educational institutions and policymakers address this divide, ensuring that every student, regardless of their internet connectivity, has an equal opportunity to learn.
The pandemic, while challenging, also presented opportunities. It accelerated the adoption of digital tools, fostered innovation in teaching methods, and highlighted the resilience and adaptability of the global student community.

As we look ahead, the lessons learned during the pandemic hold profound implications for future research and practices. There is a need to delve deeper into the long-term effects of remote learning on student performance and well-being. Exploring the potential of emerging technologies, such as augmented and virtual reality in online education, could also be a promising avenue. Furthermore, studies focusing on teacher training for digital education and the development of more inclusive online platforms, catering to students with disabilities, would be invaluable. The pandemic has also underscored the importance of mental health in academic settings, paving the way for research on online support systems for students and educators alike.

In closing, the COVID-19 pandemic, with all its disruptions, has left an indelible mark on the annals of education. This retrospective study, serves as both a reflection on a challenging past and a roadmap for a promising future, with ample avenues for exploration and growth.

\section*{Online supplementary materials}

The replication package is openly available under a CC-BY 4.0 license at the following DOI: 10.5281/zenodo.8268924. 

 \section*{Acknowledgments}
This work was supported by the Carlsberg Foundation under grant agreement number CF20-0322 (PanTra --- Pandemic Transformation).

During the preparation of this work the author(s) used ChatGPT-4 in order to ensure linguistic accuracy and enhancing the readability of this article. After using this tool/service, the author(s) reviewed and edited the content as needed and take(s) full responsibility for the content of the publication.


%%

\bibliographystyle{unsrt}  
\bibliography{bib}





% \end{thebibliography}

\appendix

\section{Appendix A (Survey Instrument)}
\label{app:Appendix}

\begin{table}[ht!]
\centering
\sisetup{
group-digits=true,
group-minimum-digits=4,
table-format=0.3,
mode=text,
detect-weight=true,
detect-family=true
}
\small
\robustify{\textbf}
\caption{Items description. Those prefixed with (*) were dropped because of their insufficient loading onto their latent variable}
\label{tab:Items}
\begin{tabular}{p{2cm} p{1cm} p{8cm} p{1cm}}
\toprule
Construct &
Item ID &
Questions &
Reference  \\
\midrule
User Satisfaction & US\_1
&
I would recommend this instructor to other students. &~\cite{eom2016determinants}
\\
& US\_2
&
I would recommend this class to other students. &~\cite{eom2016determinants}
\\
& US\_3
&
I would take a class at this university again in the future. &~\cite{eom2016determinants}
\\
& US\_4
&
I was very satisfied with this online class. &~\cite{eom2016determinants}
\\
\addlinespace
Learning Outcomes & LO\_1
&
The academic quality of this online class is on par with face-to-face classes I’ve taken. &~\cite{eom2016determinants}
\\
& LO\_2
&
I have learned as much from this online class as I might have from a face-to-face version of the course. &~\cite{eom2016determinants}
\\
& LO\_3
&
I learn more in online classes than in face-to-face classes. &~\cite{eom2016determinants}
\\
& LO\_4
&
The quality of the learning experience in online classes is better than in face-to-face classes. &~\cite{eom2016determinants}
\\
\addlinespace
Fostering Relationships & FR\_1
&
The instructor stressed the importance of fostering relationships as an effective practice for teaching. &~\cite{bailey2009effective}
\\
& FR\_2
&
The instructor showed to students empathy and passion for teaching. &~\cite{bailey2009effective}
\\
& FR\_3
&
The instructor showed a strong desire to help students be successful at the university level. &~\cite{bailey2009effective}
\\
\addlinespace
Engagement & EN\_1
&
The instructor was very engaged with the class. &~\cite{bailey2009effective}
\\
& EN\_2
&
(*) We actively used education tools (e.g. e-mails, class discussion boards, sharing student biographies, and student group projects). &~\cite{bailey2009effective}
\\
& EN\_3
&
(*) Discussions with the instructor were an important part of the online course. &~\cite{bailey2009effective}
\\
\addlinespace
Timeliness & TI\_1
&
Timeliness was taken very seriously by the instructor. &~\cite{bailey2009effective}
\\
& TI\_2
&
The instructor returned graded assignments promptly, frequently checked e-mails and responded to questions fast. &~\cite{bailey2009effective}
\\
& TI\_3
&
I had the impression the instructor checked her/his Email several times a day. &~\cite{bailey2009effective}
\\
\addlinespace
Communication & CO\_1
&
The instructor was willing to communicate with us, caring about the students. &~\cite{bailey2009effective}
\\
& CO\_2
&
The instructor used a wording that was always very clear to me and I never took anything the wrong way. &~\cite{bailey2009effective}
\\
& CO\_3
&
I had a very good communication with my instructor. &~\cite{bailey2009effective}
\\
\addlinespace
Organization & OR\_1
&
(*) It was easy to navigate the online university course website. &~\cite{bailey2009effective}
\\
& OR\_2
&
I had a full understanding of what was required to be successful in the online classes. &~\cite{bailey2009effective}
\\
& OR\_3
&
The instructor was very effective in organising the whole online course. &~\cite{bailey2009effective}
\\
\addlinespace
Technology & TEC\_1
&
The online platform where the course was run (e.g., Moodle, MS Teams, Zoom) was very effective. &~\cite{bailey2009effective}
\\
& TEC\_2
&
The online platform was able to substitute effectively the in-person physical experience. &~\cite{bailey2009effective}
\\
& TEC\_3
&
The different online tools chosen by the instructor (e.g., quizzes, wikis, chats) were NOT confusing. &~\cite{bailey2009effective}
\\
\addlinespace
Flexibility & FLEX\_1
&
The instructor was very patient with us students. &~\cite{bailey2009effective}
\\
& FLEX\_2
&
The instructor showed a high level of flexibility when new issues arose. &~\cite{bailey2009effective}
\\
& FLEX\_3
&
When some issues happened (i.e., of technological nature), the instructor was able to handle them and adapt to the new situation. &~\cite{bailey2009effective}
\\
\addlinespace
Expectations & EX\_1
&
The course goals and learning objectives were clearly established. &~\cite{bailey2009effective}
\\
& EX\_2
&
To me, it was clear from the very beginning what was expected from me. &~\cite{bailey2009effective}
\\
& EX\_3
&
The instructor performed the course as I expected. &~\cite{bailey2009effective}
\\
\addlinespace
\bottomrule
\end{tabular}
\end{table}

\end{document}